\documentclass[10pt]{iopart}
\usepackage{iopams}
\usepackage{times}

\def\pd#1#2{\frac{\partial#1}{\partial#2}}
\def\pdd#1#2#3{\ifx#2#3\frac{\partial^2#1}{\partial#2^2}\else
  \frac{\partial^2#1}{\partial#2\mkern1mu\partial#3}\fi}

\def\eqnitem#1. {\par\medbreak\leavevmode#1$^\circ$.\enspace}
\def\Example#1.{{\sl Example\hskip.5em\relax#1\unskip.}}
\def\Remark#1.{{\sl Remark\hskip.5em\relax#1\unskip.}}
\def\const{\mbox{\rm const}}
\begin{document}

\title[The Crocco transformation: order reduction and
new integrable equations]
{The Crocco transformation: order reduction and construction of
B\"acklund transformations and new integrable equations}

\author{A D Polyanin$^1$ and A I Zhurov$^{1,2}$}
\address{$^1$ Institute for Problems in Mechanics,
  Russian Academy of Sciences,\\
  101 Vernadsky Avenue, bldg 1, 119526 Moscow, Russia.}
\address{$^2$ School of Dentistry, Cardiff University,
  Heath Park, Cardiff CF14 4XY, UK.}
\ead{polyanin@ipmnet.ru}

\begin{abstract}
Wide classes of nonlinear mathematical physics equations are described that
admit order reduction through the use of the Crocco transformation, with a
first-order partial derivative taken as a new independent variable and a
second-order partial derivative taken as the new dependent variable. Associated
B\"acklund transformations are constructed for evolution equations of general
form (special cases of which are Burgers, Korteweg--de Vries, and many other
nonlinear equations of mathematical physics). The results obtained are used for
order reduction and constructing exact solutions of hydrodynamics equations
(Navier--Stokes, Euler, and boundary layer). A number of new integrable nonlinear
equations, inclusive of the generalized Calogero equation, are considered.
\end{abstract}


\maketitle

\section{Preliminary remarks}

The Crocco transformation is used in hydrodynamics for reducing the order of
the plane boundary-layer equations [1--3]. It is a transformation in which a
first-order partial derivative taken as a new independent variable and a
second-order partial derivative taken as the new dependent variable (this
applies to the equation for the stream function). So far, using the Crocco
transformation has been limited solely to the theory of boundary layer.

The present paper reveals that the domain of application of the
Crocco transformation is much broader. It can be successfully used for
reducing the order of wide classes of nonlinear equations with mixed
derivatives and constructing B\"acklund transformations for evolution equations
of arbitrary order and quite general form, special cases of which
include Burgers and Korteweg--de Vries type equations as well as many other
nonlinear equations of mathematical physics. The B\"acklund transformations
obtained with the Crocco transformation may, in turn, be used for constructing
new integrable nonlinear equations. Examples of the generalized Calogero
equation and a number of other integrable nonlinear second-, third-, and
fourth-order equations are considered. A generalization of the Crocco
transformation to the case of three independent variables is given.

It is noteworthy that various B\"acklund transformations and their applications
to specific equations of mathematical physics can be found, for example,
in~[3--14].

In the present paper, the term {\it integrable
equation\/} applies to nonlinear partial differential equations that admit
solution in terms of quadratures or solutions to linear differential or
linear integral equations.

\section{Nonlinear equations that admit order reduction with the Crocco transformation}

Consider the $n$th-order nonlinear equation with a mixed derivative
\begin{equation}
\pdd utx+[a(t)u+b(t)x]\pdd uxx=
F\left(t,\pd ux,\pdd uxx, \pd{^3u}{x^3},\dots, \pd{^nu}{x^n}\right).
\label{(2.1)}
\end{equation}

\eqnitem 1.
General property:
if $\widetilde u(t,x)$ is a solution to equation \eref{(2.1)}, then the function
\begin{equation}
u=\widetilde u(t,x+\varphi(t))+\frac 1{a(t)}[b(t)\varphi(t)-\varphi'_t(t)],\quad
a(t)\not\equiv0,
\label{(2.2)}
\end{equation}
where $\varphi(t)$ is an arbitrary function, is also a solution to equation
\eref{(2.1)}. If $a(t)\equiv0$, then $u=\widetilde u(t,x)+\varphi(t)$ is
another solution to~\eref{(2.1)}.

\eqnitem 2.
Denote
\begin{equation}
\eta=\pd ux,\quad \Phi=\pdd uxx.
\label{(2.3)}
\end{equation}
Dividing \eref{(2.1)} by $u_{xx}=\Phi$, differentiating
with respect to~$x$, and taking into account \eref{(2.3)}, we obtain
\begin{equation}
\frac{\Phi_t}\Phi-\frac{u_{tx}\Phi_x}{\Phi^2}+a(t)\eta+b(t)=
\pd{}x\frac {F(t,\eta,\Phi, \Phi_x,\dots, \Phi_x^{(n-2)})}\Phi.
\label{(2.4)}
\end{equation}

Let us pass in \eref{(2.4)} from the old variables to the Crocco variables:
\begin{equation}
t, \ x, \  u=u(t,x)\quad \ \Longrightarrow\quad \
t, \ \eta, \ \Phi=\Phi(t,\eta),
\label{(2.5)}
\end{equation}
where $\eta$ and $\Phi$ are defined by \eref{(2.3)}.
The derivatives are transformed as follows:
\[
\pd{}x=\pd\eta x\pd{}\eta=u_{xx}\pd{}\eta=\Phi\pd{}\eta,\quad \
\pd{}t=\pd{}t+\pd\eta t\pd{}\eta=\pd{}t+u_{tx}\pd{}\eta.
\]
As a result, equation \eref{(2.4)}, and hence the original equation \eref{(2.1)},
is reduced to the $(n-1)$st-order equation
\begin{equation}
\frac{a(t)\eta+b(t)}\Phi-\pd{}t\frac 1\Phi=\pd{}\eta
\left[\frac 1\Phi F\left(t,\eta,\Phi,\Phi\pd\Phi\eta,\dots,\pd{^{n-2}\Phi}{x^{n-2}}\right)\right].
\label{(2.6)}
\end{equation}
The higher derivatives are calculated by the formulas
\[
\pd{^{k} u}{x^{k}}=\pd{^{k-2}\Phi}{x^{k-2}}=
\Phi\pd{}\eta\pd{^{k-3}\Phi}{x^{k-3}},\quad
\pd{}x=\Phi\pd{}\eta,\quad k=3,\dots,n.
\]

Given a solution to the original equation (\ref{(2.1)}), formulas (\ref{(2.3)})
define a solution to equation~\eref{(2.6)} in parametric form.

Let $\Phi=\Phi(t,\eta)$ be a solution to equation~\eref{(2.6)}.
Then, in view of \eref{(2.3)}, the function $u(t,x)$ satisfies
the equation
\begin{equation}
u_{xx}=\Phi(t,u_x),
\label{(2.7)}
\end{equation}
which can be treated as an ordinary differential equation in~$x$
with parameter~$t$. The general solution to equation \eref{(2.7)}
may be written in parametric form as
\begin{equation}
x=\int^\eta_{\eta_0}\frac{\rmd s}{\Phi(t,s)}+\varphi(t),\quad
u=\int^\eta_{\eta_0}\frac{s\,\rmd s}{\Phi(t,s)}+\psi(t),
\label{(2.8)}
\end{equation}
where $\varphi(t)$ ¨ $\psi(t)$ are arbitrary functions and $\eta_0$ is an
arbitrary constant. Since the derivation of~\eref{(2.6)} is based on
differentiating~\eref{(2.1)}, one of the arbitrary functions in
solution~\eref{(2.8)} is redundant. In order to remove this redundancy, it
suffices to substitute \eref{(2.8)} into~\eref{(2.1)}. However, it more
convenient to take advantage of the solution property \eref{(2.2)} and note
that solution \eref{(2.8)} must also possess this property. In view of
this, the general solution to the original equation~\eref{(2.1)} can be
rewritten in the parametric form
\begin{equation}
x=\int^\eta_{\eta_0}\frac{\rmd s}{\Phi(t,s)}+\varphi(t),\quad
u=\int^\eta_{\eta_0}\frac{s\,\rmd s}{\Phi(t,s)}+
\frac 1{a(t)}[b(t)\varphi(t)-\varphi'_t(t)],
\label{(2.9)}
\end{equation}
where $\varphi(t)$ is an arbitrary function.

\medskip
\Example 1 (generalized Calogero equation).
With $F=f(t,u_x)u_{xx}+g(t,u_x)$, which
corresponds to the nonlinear second-order equation
\begin{equation}
u_{tx}=[f(t,u_x)-a(t)u-b(t)x]u_{xx}+g(t,u_x),
\label{(2.10)}
\end{equation}
passing to the Crocco variables \eref{(2.5)}, \eref{(2.3)}
leads to the first-order equation
\[
\frac{a(t)\eta+b(t)}\Phi-\pd{}t\frac 1\Phi=
\pd{}\eta\left[f(t,\eta)+\frac {g(t,\eta)}\Phi\right],
\]
which becomes linear with the substitution $\Phi=1/\Psi$.

In the special case of $a(t)=-1$, $b(t)=0$, $f(t,u_x)=0$, and
$g(t,u_x)=g(u_x)$, equation \eref{(2.10)} reduces to the Calogero equation,
which was considered in [15,~16] (see also [3,~p.~433--434]).

\Example 2 (equation arising in gravitation theory).
The nonlinear third-order equation
\[
u_{txx}=kuu_{xxx},
\]
which is cross-disciplinary between projective geometry and gravitation theory
[16, 17], can be reduced, by integrating with respect to~$x$, to the
form
\begin{equation}
u_{tx}=kuu_{xx}-\case12 kw_x^2+\psi(t),
\label{(2.11)}
\end{equation}
where $\psi(t)$ is an arbitrary function. \Eref{(2.11)} is a special case
of equation \eref{(2.10)}, and hence can be reduced to
a linear first-order equation.

\Example 3 (Navier--Stokes and Euler equations).
An unsteady three-dimensional flow of a viscous incompressible fluid
may be described by the Navier--Stokes and continuity equations
\begin{equation}\eqalign{
\pd {V_n}t&+V_1\pd {V_n}x+V_2\pd {V_n}y+V_3\pd {V_n}z\cr
&=-\frac 1{\rho}\nabla_n P
+\nu\left(\pdd {V_n}xx+\pdd {V_n}yy+\pdd {V_n}zz\right),\quad n=1,\,2,\,3,\cr
\pd {V_1}x&+\pd {V_2}y+\pd {V_3}z=0,\cr}
\label{(2.12)}
\end{equation}
where $x$,~$y$, and~$z$ are Cartesian coordinates, $t$ time, $V_1$,~$V_2$,
and~$V_3$ the fluid velocity components, $P$ pressure, and $\rho$ the fluid
density; also $\nabla_1 P=\partial P/\partial x$, $\nabla_2 P=\partial
P/\partial y$, and $\nabla_3 P=\partial P/\partial z$. Equations
\eref{(2.12)} are obtained under the assumption that the bulk forces are
potential and included into pressure. In the degenerate case of $\nu=0$,
equations~\eref{(2.12)} become the Euler equations for an ideal (inviscid)
fluid.

The equations of motion of a viscous incompressible fluid, \eref{(2.12)},
admit exact three-dimensional solutions of the form
\[
\eqalign{
&\,V_1=u,\quad V_2=-\frac12y\pd ux,\quad V_3=-\frac12z\pd ux,\cr
&\frac P\rho=\frac14p(t)(y^2+z^2)+s(t)-\frac12 u^2
+\nu\pd ux-\int \pd ut\,\rmd x,\cr}
\]
where $p(t)$ and $s(t)$ are arbitrary functions of time~$t$,
and $u=u(t,x)$ satisfies the nonlinear third-order equation
\begin{equation}
\pdd utx+u\pdd uxx-\frac12\left(\pd ux\right)^{\!2}=
\nu \frac{\partial^3u}{\partial x^3}+p(t),
\label{(2.13)}
\end{equation}
which is a special case of equation \eref{(2.1)} with $a(t)=1$, $b(t)=0$, and
$F=\nu u_{xxx}+\frac12u_x^2+p(t)$.

The Crocco transformation \eref{(2.5)} brings \eref{(2.13)} to
the nonlinear second-order equation
\begin{equation}
\pd\Phi t+[\case12\eta^2+p(t)]\pd\Phi\eta=\nu\Phi^2\pdd\Phi\eta\eta,
\label{(2.14)}
\end{equation}
which can be rewritten in the form of a nonlinear equation of convective
thermal conduction with a parabolic, Poiseuille-type velocity profile:
\[
\pd\Psi t+[\case12\eta^2+p(t)]\pd\Psi\eta=
\nu\pd{}\eta\left(\frac 1{\Phi^2}\pd\Psi\eta\right),\quad \
\Psi=\frac 1\Phi.
\]

It should be noted that in the special case of inviscid fluid ($\nu=0$),
the original nonlinear equation \eref{(2.13)} is reducible to the
linear first-order partial differential equation \eref{(2.14)}, which
can be solved by the method of characteristics.

\Example 4 (system of hydrodynamic-type equations).
Consider the system of equations
\begin{eqnarray}
&\pdd utx+u\pdd uxx-\left(\pd ux\right)^{\!2}&=
  \nu \frac{\partial^3u}{\partial x^3}+q(t)\pd ux+p(t),\label{(2.15)}\\
&\pd v t+u\pd v x-v \pd ux&=\nu\pdd v xx,\label{(2.16)}
\end{eqnarray}
which describes several classes of exact solutions to the Navier--Stokes
equations in two and three dimensions [3, 18--20]. The nonlinear equation
\eref{(2.15)} is independent of~$v$ and can be treated separately. Although
linear in~$v$, equation \eref{(2.16)} involves the function~$u$, which is
governed by equation~\eref{(2.15)}.

The Crocco transformation \eref{(2.5)} brings system
\eref{(2.15)}--\eref{(2.16)} to the form
\begin{eqnarray}
\pd\Phi t+(\eta^2+q\eta+p)\pd\Phi\eta&=
  (\eta+q)\Phi+\nu\Phi^2\pdd\Phi\eta\eta,\label{(2.17)}\\
\pd v t+(\eta^2+q\eta+p)\pd v \eta&=
  \eta v+\nu\Phi^2\pdd v \eta\eta.\label{(2.18)}
\end{eqnarray}
Here and henceforth, the arguments of $p(t)$ and~$q(t)$ are omitted for brevity.
Equation~\eref{(2.18)} was obtained using the representation of the mixed derivative $u_{tx}$
obtained from~\eref{(2.15)}.

\Eref{(2.18)} has exact solutions of the form
\begin{equation}
v=A\eta + B\Phi+C,
\label{(2.19)}
\end{equation}
where $A=A(t)$, $B=B(t)$, and $C=C(t)$ are unknown functions
determined from an appropriate system of ordinary differential
equations. This fact can be proved by substituting
\eref{(2.19)} into \eref{(2.18)} and taking into account~\eref{(2.17)}.

Formula \eref{(2.19)} allows one to arrive at the following important result
with regard to solutions of the original equation~\eref{(2.16)}.
Let $u=u(t,x)$ be a solution to equation~\eref{(2.15)}.
Then equation \eref{(2.16)} admits the solution
\begin{equation}
 v=A'_t+Aq+A\pd ux+B\pdd uxx,
\label{(2.20)}
\end{equation}
where $A=A(t)$ and $B=B(t)$ satisfy the ordinary differential equations
\begin{eqnarray}
A''_{tt}+qA'_t+(p+q'_t)A&=0,\label{(2.21)}\\
B'_t+qB&=0.\label{(2.22)}
\end{eqnarray}
The general solution to \eref{(2.22)} is
$\displaystyle B=C_1\exp\left(-\int q\,\rmd t\right)$,
where $C_1$ is an arbitrary constant.

Listed below are some exact solutions to equation \eref{(2.15)}
representable in terms of elementary functions and suitable for finding exact
solutions to equation~\eref{(2.16)} using formulas~\eref{(2.20)}.

\eqnitem 1.  Generalized separable solution rational in $x$:
\[
u=-\alpha '_t(t)+\beta (t)[x+\alpha (t)]-\frac{6\nu}{x+\alpha (t)},\quad
q=-4\beta ,\quad p=\beta '_t+3\beta ^2,
\]
where $\alpha =\alpha (t)$ and $\beta =\beta (t)$ are arbitrary functions.

\eqnitem 2. Generalized separable solution exponential in~$x$:
\[
u=\alpha (t)\rme^{-\sigma x}+\beta (t),\quad
p=0,\quad
q=\frac {\alpha '_t}\alpha -\sigma \beta -\sigma ^2\nu,
\]
where $\alpha =\alpha (t)$ and $\beta =\beta (t)$ are arbitrary functions
and $\sigma$ is an arbitrary constant. By choosing periodic functions as
$\alpha (t)$ and $\beta (t)$, one obtains time-periodic solutions.

\eqnitem 3. Multiplicative separable solution periodic in~$x$:
\[
\eqalign{
u&=\alpha (t)\sin(\sigma x+C_1),\quad
\alpha (t)=C_2\exp\left[-\nu \sigma ^2t+\int q(t)\,\rmd t\right],\cr
p&=-\sigma ^2\alpha ^2(t),\quad
\hbox{$q=q(t)$ is an arbitrary function},\cr}
\]
where $C_1$, $C_2$, and $\sigma $ are arbitrary constants. By setting
$q(t)=\nu \sigma ^2+\varphi'_t(t)$ with periodic $\varphi(t)$, one obtains
a periodic solution in both $x$ and~$t$.

More complicated solutions to equation \eref{(2.15)} can be found in~[20].

\section{Some generalizations}

Consider the nonlinear $n$th-order equation
\begin{equation}\fl
c(t)u_{tx}+[a(t)u+b(t)x]u_{xx}+d(t)(u_xu_{tx}-u_tu_{xx})
=F(t,u_x,u_{xx},\ldots,u_x^{(n)}),
\label{(3.1)}
\end{equation}
which becomes \eref{(2.1)} for $c(t)=1$ and $d(t)=0$.

\eqnitem 1.
General property:
if $\widetilde u(t,x)$ is a solution to equation~\eref{(3.1)}, then the function
\[
u=\widetilde u(t,\,x+\varphi(t))+\psi(t),
\]
where $\varphi=\varphi(t)$ and $\psi=\psi(t)$ are related by
$d(t)\psi'_t-a(t)\psi=c(t)\varphi'_t-b(t)\varphi$ (either function
can be chosen arbitrarily), is also a solution to~\eref{(3.1)}.

\eqnitem 2.
Let us divide \eref{(3.1)} by $u_{xx}$, differentiate
the resulting equation with respect to~$x$, and then
pass to the Crocco variables \eref{(2.5)}, \eref{(2.3)}
to obtain the $(n-1)$st-order equation
\[\fl
\frac{a(t)\eta+b(t)}\Phi-[d(t)\eta+c(t)]\pd{}t\frac 1\Phi=
\pd{}\eta\left[\frac 1\Phi F\left(t,\eta,\Phi,\Phi\pd\Phi\eta,
\dots,\pd{^{n-2}\Phi}{x^{n-2}}\right)\right].
\]

\Example. In the special case of $n=3$, $a(t)=b(t)=c(t)=0$, $d(t)=1$, and
$F=[f(u_{xx})]_x$, \eref{(3.1)} is a general boundary layer equation for a
non-Newtonian fluid~[3], with $u$ being the stream function. By the Crocco
transformation~\eref{(2.5)}, this equation can be reduced to the second-order
equation $\eta\Phi_t=\Phi^2[f(\Phi)]_{\eta\eta}$, which can be
linearized by the substitution $\Psi=1/\Phi$ if $f(\Phi)=1/\Phi$.

\Remark. \Eref{(3.1)} can be generalized by adding
the arguments $J_x$,~\dots,~$J_x^{(m)}$,
with $J=u_{xx}u_{txx}-u_{tx}u_{xxx}$, to the function~$F$.

\section{Using the Crocco transformation for constructing RF-pairs and
B\"acklund transformations}

Consider a fairly general $n$th-order evolution equation
\begin{equation}
u_t+[a(t)u+b(t)x]u_x=F(t,u_x,u_{xx},u_{xxx},\dots,u^{(n)}_x).
\label{(4.1)}
\end{equation}

General property: if $\widetilde u(t,x)$ is a solution to
equation~\eref{(4.1)}, then the function
\[
u=\widetilde u(t,x+\psi(t))+C,
\]
where $C$ is an arbitrary constant and $\psi=\psi(t)$ satisfies the linear
ordinary differential equation $\psi'_t-b(t)\psi+Ca(t)=0$, is
also a solution to~\eref{(4.1)}.

Differentiating \eref{(4.1)} with respect to $x$ yields an $(n+1)$st-order
equation with a mixed derivative of the form~\eref{(2.1)}:
\begin{equation}
\fl u_{tx}+[a(t)u+b(t)x]u_{xx}
 =-a(t)u_x^2-b(t)u_x+
\pd{}xF(t,u_x,u_{xx},u_{xxx},\dots,u^{(n)}_x).
\label{(4.2)}
\end{equation}
By passing in \eref{(4.2)} from $t$, $x$, $u$ to
the Crocco variables \eref{(2.5)}, we one arrives at the $n$th-order equation
\begin{equation}
\fl\frac{3a(t)\eta+2b(t)}\Phi-\pd{}t\frac 1\Phi
+[a(t)\eta^2+b(t)\eta]\pd{}\eta\frac 1\Phi=
\pd{^2}{\eta^2}
F\left(t,\eta,\Phi,\Phi\pd\Phi\eta,\dots,\pd{^{n-2}\Phi}{x^{n-2}}\right).
\label{(4.3)}
\end{equation}

Equations \eref{(4.1)} and \eref{(4.3)} are linked by the
B\"acklund transformation
\begin{equation}
\eqalign{
&u_t+[a(t)u+b(t)x]\eta=F(t,\eta,\Phi,\Phi_x,\dots,\Phi^{(n-2)}_x),\cr
&u_x=\eta,\quad u_{xx}=\Phi.\cr}
\label{(4.4)}
\end{equation}

\Remark.
Sometimes, it is convenient to rewrite \eref{(4.3)}
in the form
\[
\Psi_t-[a(t)\eta^2+b(t)\eta]\Psi_\eta-[3a(t)\eta+2b(t)]\Psi=
-\pd{^2}{\eta^2}F, \quad
\Psi=\frac 1\Phi.
\]

\Example 1. The unnormalized Burgers equation
\begin{equation}
u_t+auu_x=\beta u_{xx}
\label{(4.5)}
\end{equation}
is a special case of \eref{(4.1)} with $a(t)=a=\const$, $b(t)=0$, and
$F=\beta u_{xx}=\beta \Phi$. By the B\"acklund transformation~\eref{(4.4)},
equation~\eref{(4.5)} can be reduced to
\[
\Phi_t-a\eta^2\Phi_\eta+3a\eta\Phi=\beta\Phi^2\Phi_{\eta\eta}.
\]

\Example 2.
The nonlinear second-order equation
\begin{equation}
u_t+[a(t)u+b(t)x]u_x=\frac {f(t,u_x)}{u_{xx}}+g(t,u_x)
\label{(4.6)}
\end{equation}
is a special case of \eref{(4.1)}. The B\"acklund transformation
\eref{(4.4)} reduces \eref{(4.6)} to the equation
\[\fl
\frac{3a(t)\eta+2b(t)}\Phi-\pd{}t\frac 1\Phi
+[a(t)\eta^2+b(t)\eta]\pd{}\eta\frac 1\Phi=
\pd{^2}{\eta^2}\left[\frac {f(t,\eta)}\Phi+g(t,\eta)\right],
\]
which becomes linear after substituting $\Phi=1/\Psi$.

\Example 3. The unnormalized Burgers Korteweg--de Vries equation
\begin{equation}
u_t+auu_x=\beta u_{xxx}
\label{(4.7)}
\end{equation}
is a special case of \eref{(4.1)} $a(t)=\const$, $b(t)=0$, and $F=\beta
u_{xxx}=\beta \Phi\Phi_\eta$. The B\"acklund transformation
\eref{(4.4)} reduces \eref{(4.7)} to the equation
\[
\Phi_t-a\eta^2\Phi_\eta+3a\eta\Phi=\beta\Phi^2(\Phi\Phi_\eta)_{\eta\eta},
\]
which, after submitting $\Phi=\theta^{1/2}$, becomes
\[
\theta_t-a\eta^2\theta_\eta+6a\eta\theta=\beta\theta^{3/2}\theta_{\eta\eta\eta}.
\]

\Example 4.
The nonlinear third-order equation
\begin{equation}
u_t+auu_x=\frac {f(t,u_x)}{u_{xx}^3}u_{xxx}
\label{(4.8)}
\end{equation}
can be reduced, using the B\"acklund transformation
\eref{(4.4)} with $b(t)\equiv 0$ and
$F=f(t,u_x)u_{xx}^{-3}u_{xxx}=f(t,\eta)\Phi^{-2}\Phi_\eta$
followed by substituting $\Phi=1/\Psi$, to the linear equation
\[
\Psi_t-a\eta^2\Psi_\eta-3a\eta\Psi=[f(t,\eta)\Psi_\eta]_{\eta\eta}.
\]

\Example 5.
The linear third-order equation
\begin{equation}
u_t=\alpha u_{xxx}+\beta u_{xx}
\label{(4.9)}
\end{equation}
is a special case of \eref{(4.1)} with $F=\alpha u_{xxx}+\beta
u_{xx}=\alpha \ Phi\Phi_\eta+\beta\Phi$, $a(t)\equiv 0$, and $b(t)\equiv 0$.
By applying to \eref{(4.9)} the B\"acklund transformation~\eref{(4.4)}
and then substituting $\Phi=1/\Psi$, one arrives at the nonlinear equation
\begin{equation}
\Psi_t=\alpha(\Psi^{-3}\Psi_\eta)_{\eta\eta}+\beta(\Psi^{-2}\Psi_\eta)_\eta.
\label{(4.10)}
\end{equation}
The special cases of \eref{(4.10)} with $\alpha=0$, $\beta\not=0$ and
$\beta=0$, $\alpha\not=0$ were copnsidered in [21] and~[3], respectively.

\Example 6. The linear fourth-order equation
$u_t=\alpha u_{xxxx}$
is reduced, using the same transformation as in the preceding
example and substituting $\Phi=\theta^{1/2}$,
to the nonlinear fourth-order equation
\[
\theta_t=\alpha\theta^{3/2}(\theta^{1/2}\theta_{\eta\eta})_{\eta\eta}.
\]

\Remark. \Eref{(4.3)} remains unchanged if the sum $p(t)u+q(t)x+s(t)$,
with arbitrary functions $p(t)$, $q(t)$, and~$s(t)$,
is added to the right-hand side of~\eref{(4.1)} and that of
the first equation in~\eref{(4.4)}.

{\it Corollary}. If equation \eref{(4.1)} is integrable for
some right-hand side~$F$, then the equation with the more complicated
right-hand side $F+p(t)u+q(t)x+s(t)$ is also integrable.

\Example 1.
Since the Burgers equation $u_t+auu_x=bu_{xx}$ is integrable,
the more complicated equation
\[
u_t+auu_x=bu_{xx}+p(t)u+q(t)x+s(t)
\]
is also integrable.

\Example 2. Likewise, since the Korteweg--de Vries equation
$u_t+auu_x=bu_{xxx}$ is integrable, the more complicated equation
\[
u_t+auu_x=bu_{xxx}+p(t)u+q(t)x+s(t)
\]
is also integrable.

\section{Extension of the Crocco transformation to the case
of three independent variables. Application to unsteady boundary-layer
equations}

Transformation \eref{(2.5)} can be extended to
the cases of more independent variables. In particular,
it can be shown that the Crocco transformation
\begin{equation}\fl
\def\quad{\ \ \ }
t, \ x, \ y, \  u=u(t,x,y)\quad \Longrightarrow\quad
t, \ x, \ \eta, \ \Phi=\Phi(t,x,\eta), \quad
\hbox{where}\quad \eta=u_y, \ \Phi=u_{yy},
\label{(5.1)}
\end{equation}
reduces the order of the $n$th-order equation
\begin{eqnarray}
\fl [a(t,x)u+b(t,x)y]u_{yy}+c_1(t,x)u_{ty}+c_2(t,x)u_{xy}+
d_1(t,x)(u_yu_{ty}-u_tu_{yy})\nonumber\\
{}+d_2(t,x)(u_yu_{xy}-u_xu_{yy})=
F(t,x,u_y,u_{yy},\ldots,u_y^{(n)}).
\label{(5.2)}
\end{eqnarray}

\Example. Consider the Prandtl system
\begin{equation}
\eqalign{
u_t+uu_x+vu_y&=\nu u_{yy}+f(t,x),\cr
u_x+v_y&=0,\cr}
\label{(5.3)}
\end{equation}
which describes a flat unsteady boundary layer with pressure gradient
($u$~and~$v$ the fluid velocity components) [1--3].
Equations~\eref{(5.3)} can be reduced,
by introducing a stream function~$w$ such that $u=w_y$ and $v=-w_x$,
to a single third-order equation [1,~3]:
\begin{equation}
w_{ty}+w_yw_{xy}-w_xw_{yy}=\nu w_{yyy}+f(t,x).
\label{(5.4)}
\end{equation}
This equation is a special case of \eref{(5.2)} (up to the obvious renaming
$u\rightleftarrows w$).

Dividing \eref{(5.4)} by $w_{yy}$ followed by differentiating with respect
to~$y$ and passing from $t$,~$x$, $y$,~$w$ to the Crocco variables
$t$,~$x$, $\eta=w_y$, $\Phi=w_{yy}$, one arrives at
the second-order equation
\begin{equation}
\pd\Phi t+\eta \pd\Phi x+f(t,x)\pd\Phi\eta=\nu\Phi^2\pdd\Phi\eta\eta,
\label{(5.5)}
\end{equation}
which is reduced, with the substitution $\Phi=1/\Psi$,
to the nonlinear heat equation
\begin{equation}
\pd\Psi t+\eta \pd\Psi x+f(t,x)\pd\Psi\eta=
\nu\pd{}\eta\left(\frac 1{\Psi^2}\pd\Psi\eta\right).
\label{(5.6)}
\end{equation}

\Remark.
In the steady-state case with $\partial/\partial t=0$ and $f(t,x)=0$,
equation~\eref{(5.5)} reduces to one considered in~[1,~3].

\eqnitem 1.
In the special case $f(t,x)=f(t)$, equation \eref{(5.6)} admits
an exact solution of the special form
\[
\Psi=Z(\xi,\tau),\quad \xi=x-\eta t+\int tf(t)\,\rmd t,\quad \tau=\frac 13t^3.
\]
Hence we arrive at the integrable equation
\begin{equation}
\pd Z\tau=\nu\pd{}\xi\left(\frac 1{Z^2}\pd Z\xi\right),
\label{(5.7)}
\end{equation}
which can be reduced to the linear heat equation [3,~21].

\eqnitem 2.
In the more general case $f(t,x)=f(t)x+g(t)$, we
have solutions of the special form
\[
\Psi=Z(\xi,\tau),\quad \xi=\varphi(t)x+\psi(t)\eta+\theta(t),\quad
\tau=\int \psi^2(t)\,\rmd t,
\]
where $\varphi=\varphi(t)$, $\psi=\psi(t)$, and $\theta=\theta(t)$
are determined by the linear system of ordinary differential equations
\[
\varphi'_t+f\psi=0,\quad
\psi'_t+\varphi=0,\quad
\theta'_t+g\psi=0.
\]
As a result, we arrive at an integrable equation \eref{(5.7)}.

\section*{Acknowledgments}
The work was carried out under partial financial support of the Russian
Foundation for Basic Research (grants No.~\hbox{08-01-00553},
No.~\hbox{08-08-00530} and No.~\hbox{09-01-00343}).

\end{document}